\documentclass[sigconf]{acmart}

\usepackage{subcaption}
\usepackage{tcolorbox}
\usepackage[english]{babel}
\usepackage{graphicx,wrapfig,lipsum}
\usepackage[utf8]{inputenc} 
\usepackage{xspace}
\usepackage{adjustbox}

\usepackage{float}
\floatstyle{plaintop}
\restylefloat{table}

\usepackage{balance}

\AtBeginDocument{%
  \providecommand\BibTeX{{%
    \normalfont B\kern-0.5em{\scshape i\kern-0.25em b}\kern-0.8em\TeX}}}

\acmDOI{xx.xxx/xxx_x}

\acmISBN{978-1-4503-8104-8/21/03}

\setcopyright{acmcopyright}
\acmConference[SAC'21]{ACM SAC Conference}{March 22-March 26, 2021}{Gwangju, South Korea}
\acmYear{2021}
\copyrightyear{2021}

\renewcommand\footnotetextcopyrightpermission[1]{} 
\acmPrice{15.00}

\begin{document}

\title{Software Engineering Meets Deep Learning: A Mapping Study}



\author{Fabio Ferreira}
\affiliation{
Center of Informatics \\ 
Federal Institute of the Southeast of Minas Gerais (IF Sudeste MG)\\
Barbacena, Brazil}
\email{fabio.ferreira@ifsudestemg.edu.br}

\author{Luciana Lourdes Silva}
\affiliation{
Department of Computing\\
Federal Institute of Minas Gerais (IFMG)\\
Ouro Branco, Brazil}
\email{luciana.lourdes.silva@ifmg.edu.br }

\author{Marco Tulio Valente}
\affiliation{
ASERG Group - Department of Computer Science - 
Federal University of Minas Gerais (UFMG)\\
Belo Horizonte, Brazil}
\email{mtov@dcc.ufmg.br}

\begin{abstract}
Deep Learning (DL) is being used nowadays in many traditional Software Engineering (SE) problems and tasks. 
However, since the renaissance of DL techniques is still very recent, we lack works that summarize and condense the most recent and relevant research conducted at the intersection of DL and SE. 
Therefore, in this paper, we describe the first results of a mapping study covering 81 papers about DL \& SE. Our results confirm that DL is gaining momentum among SE researchers over the years and that the top-3 research problems tackled by the analyzed papers are documentation, defect prediction, and testing.

\end{abstract}




\maketitle

\section{Introduction}
\label{sec:intro}
Deep Learning (DL) applications are increasingly crucial in many areas, such as automatic text translation~\cite{googleTranslate2016}, image recognition~\cite{Kaiming:2015},
self-driving cars
\cite{Huval:2015}, and smart cities\cite{Mohammadi:2018}.
Moreover, various frameworks are available nowadays to facilitate the implementation of DL applications, such as TensorFlow\footnote{https://www.tensorflow.org} and PyTorch\footnote{https://pytorch.org}. Recently, Software Engineering (SE) researchers are also starting to explore the application of DL in typical SE problems, such documentation~\cite{p62, p1, p4}, defect prediction~\cite{p82, p36,p15,p42}, and testing~\cite{p11,p14,p39}.


However, since the cross-pollination between DL \& SE is recent, 
we do not have a clear map of the research conducted by combining these two areas. This map can help other researchers interested in starting to work on the application of DL in SE. It can also help researchers that already work with DL \& SE to have a clear picture of similar research in the area. Finally, mapping the research conducted in the intersection of DL \& SE might help practitioners and industrial organizations better understand the problems, solutions, and opportunities in this area.

In this paper, we provide the results of an effort to review and summarize the most recent and relevant literature about DL \& SE. To this purpose, we collect and analyze 81 papers recently published in major SE conferences and journals. We show the growth of the number of papers on DL \& SE over the years. We also reveal the most common recent problems tackled by such papers and provide data on the most common DL techniques used by SE researchers. Next, we highlight papers that achieved the same results using only traditional Machine Learning-based techniques. Finally, we discuss the main drawbacks and strengths of using DL techniques to solve SE-related problems.


\section{Deep Learning in a Nutshell}
DL is a subfield of Machine Learning (ML) that relies on multiple layers of Neural Networks (NN) to model high-level representations~\cite{goodfellow2016deep}. Similar to traditional ML, DL techniques are suitable for classification, clustering, and regression problems. 
The key diffe\-rence between traditional ML and DL techniques is that while in traditional ML approaches the features are handcrafted, in DL they are selected by neural networks automatically~\cite{lecun2015deep}.
Currently, there are many types of NNs~\cite{goodfellow2016deep} and below, we outline four common classes of NNs that are useful in SE problems:
\vspace{0.15cm}  

\noindent {\em Multilayer Perceptrons (MLP):} They are suitable for classification and regression prediction problems and can be used with different types of data, such as image, text, and time series data. Besides, when evaluating the performance of different algorithms on a particular problem, we can use MLP results as a baseline of comparison. Basically, MLPs consist of one or more layers of neurons. The input layer receives the data, the hidden layers provide abstraction levels, and the output layer is responsible for making predictions.
\vspace{0.15cm}



\noindent {\em Convolutional Neural Networks (CNN):} Although they were designed for image recognition, we can use CNN for other classification and regression problems. They also can be adapted to different types of data, such as text and sequences of input data. In summary, the input layer in CNNs receives the data, and the hidden layers are responsible for feature extraction. There are three types of layers in CNNs, such as convolution layers (which filter an input multiple times to build a feature map), pooling layers (responsible for reducing the spatial size of the feature map), and fully-connected layers. Then, the CNN output can feed a fully connected layer to create the model. 

\vspace{0.15cm}

\noindent {\em Recurrent Neural Networks (RNN):} They are a specialized type of NN for sequence prediction problems, i.e., they are designed to receive historical sequence data and predict the next output value(s) in the sequence. The main difference regarding the traditional MLP can be thought as loops on the MLP architecture. The hidden layers do not use only the current input but also the previously received inputs. Conceptually, this feedback loop adds memory to the network. The Long Short-Term Memory (LSTM) is a particular type of RNN able to learn long-term dependencies. LSTM is one of the most used RNNs in many different applications with outstanding results~\cite{Wang:2017}.
\vspace{0.15cm}

\noindent {\em Hybrid Neural Network Architectures (HNN):} They refer to architectures using two or more types of NNs. Usually, CNNs and RNNs are used as layers in a wider model. As an example from the industry, Google's translate service uses LSTM RNN
architectures~\cite{googleTranslate2016}. 

\section{Study Design}
\label{sec:design}
We conducted the mapping study following four process steps: (1) definition of research questions, (2) search process, (3) selection of studies, and (4) quality evaluation.

\subsection{Research Questions}
This study aims to identify and analyze DL techniques from the purpose of understanding their application in the context of software engineering. To achieve this goal, we established three research questions:

\begin{itemize}
\item {\textbf{(RQ1)}  \textit{What SE problems are solved by DL?}}
\item {\textbf{(RQ2)}  \textit{What DL techniques are used in SE problems?}}
\item {\textbf{(RQ3)}  \textit{How does DL compare with other ML techniques used in SE problems?}}
\end{itemize}

In RQ1, the expected result is a list of primary studies categorized by SE research problem. In RQ2, we expect to identify the most common DL techniques used by the analyzed papers. Finally, in RQ3, we pretend to clarify whether DL techniques necessarily improve the results of solving SE problems comparable with traditional ML approaches.

\subsection{Search process}
To collect the papers, we used the search string {\em ``deep learn*''} and applied this query to titles, keywords, and abstracts of articles in the following digital libraries: Scopus, ACM Digital Library, IEEE Xplore, Web of Science, SpringerLink, and Wiley Online Library. However, we only considered papers published in SE conferences and journals indexed by CSIndexbr~\cite{csindexbr}, which is a Computer Science Index system.\footnote{https://csindexbr.org} CSIndexbr is considered a GOTO ranking~\cite{gotorankings}, i.e., an information system that provides Good, Open, Transparent, and Objective data about CS departments and institutions.\footnote{http://gotorankings.org}

The SE venues listed by CSIndexbr are presented in Table~\ref{tab:venues}. As can be observed, the
system indexes 15 major conferences and 12 major journals in SE, including 
top-conferences (ICSE, FSE, and ASE),  top-journals (IEEE TSE
and ACM TOSEM) and also next-tier conferences (MSR, ICSME,
ISSTA, etc) and journals  (EMSE, JSS, IST, etc). CSIndexbr follows
a quantitative criteria, based on h5-index, number
of papers submitted and accepted to index the venues.  
\begin{table}[!t]
 \caption{Venues}
 \vspace{-0.3cm}
 \centering 
 \begin{adjustbox}{width=.47\textwidth}
 \begin{tabular}{l l r} 
 \toprule
 \multicolumn{1}{l}{\bf Acronym} & 
 \multicolumn{1}{l}{\bf Name} \\
 \midrule
 ICSE & Int.~Conference on Software Engineering \\ 
 FSE & Foundations of Software Engineering \\ 
 MSR & Mining Software Repositories\\ 
 ASE & Automated Software Engineering\\ 
 ISSTA & Int.~Symposium on Software Testing and Analysis\\
 ICSME & Int.~Conf. on Software Maintenance and Evolution\\ 
 ICST & Int.~Conf. on Software Testing, Validation and Verification\\ 
 MODELS & Int.~Conf. on Model Driven Engineering Languages and Systems \\ 
 SANER & Int.~Conf. on Software Analysis, Evolution and Reengineering\\ 
 SLPC & Systems and Software Product Line Conference\\ 
 RE & Int.~Requirements Engineering Conference\\ 
 FASE & Fundamental Approaches to Software Engineering\\ 
 ICPC & Int.~Conf. on Program Comprehension\\ 
 ESEM & Int.~Symp. on Empirical Software Engineering and Measurement\\ 
 ICSA & Int.~Conference on Software Architecture\\ 
 \midrule
 IEEE TSE & IEEE Transactions on  Software Engineering\\
 ACM TOSEM & ACM Transactions on Software Engineering and Methodology\\
 JSS & Journal of Systems and Software\\
 IEEE Software & IEEE Software\\
 EMSE & Empirical Software Engineering\\
 SoSyM & Software and Systems Modeling\\
 IST & Information and Software Technology\\
 SCP & Science of Computer Programming\\
 SPE & Software Practice and Experience\\
 SQJ & Software Quality Journal\\
 JSEP & Journal of Software Evolution and Process\\
 REJ & Requirements Engineering Journal\\
 \bottomrule
 \end{tabular}
 \end{adjustbox}
 \label{tab:venues}
 \vspace{-0.5cm}
\end{table}

By searching for {\em ``deep learn*''} from August/2019 to December/2019 we 
found 141 distinct papers in the conferences and journals listed in Table~\ref{tab:venues}. 

\subsection{Study selection}
In this step, we filtered the studies retrieved from the search process to exclude papers not aligned with the study goals. First, we removed papers with less than 10 pages, due to our decision to focus on full papers only. The only exception are papers published at IEEE Software (magazine). In this case, we defined a threshold of six pages to select the papers. After applying this size threshold, we eliminated 49 papers. 

\begin{figure}[!ht]
  \centering
  \includegraphics[scale=1.0,width=0.47\textwidth]{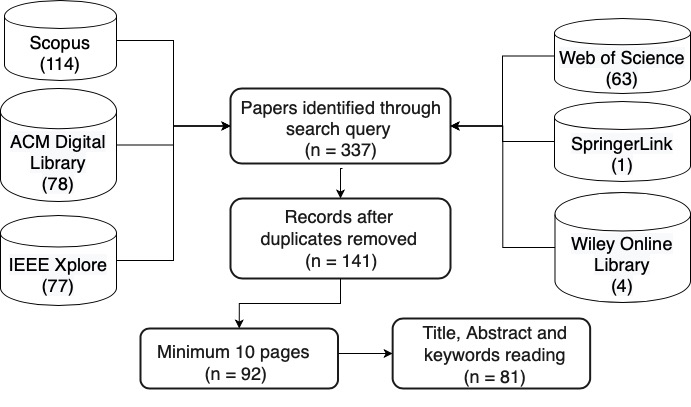}
  \caption{Steps for selecting papers}
  \label{fig:methodology}
\end{figure}

Then, we carefully
read the title and abstract of the remaining papers to confirm they indeed qualify as research that uses DL on SE-related problems.
We eliminated 11 papers, including 5 papers that are not related to SE (e.g.,~one paper that evaluates an ``achievement-driven methodology to give students more control of their learning with enough flexibility to engage them in {\em deeper learning}'')
 two papers published
in other tracks (one paper at ICSE-SEET and one
paper at ICSE-SEIP), two papers that only mention
DL in the abstract, and two papers 
that were supersed by a journal version, i.e., we discarded
the conference version and considered the extended version of the
work. Our final dataset has {\bf 81 papers}. Figure~\ref{fig:methodology} summarizes the steps we followed for selecting the papers.

\section{Results}
\label{sec:results}


\subsection{Overview}

\begin{figure}[!ht]
  \centering
  \includegraphics[scale=1.0,width=0.37\textwidth]{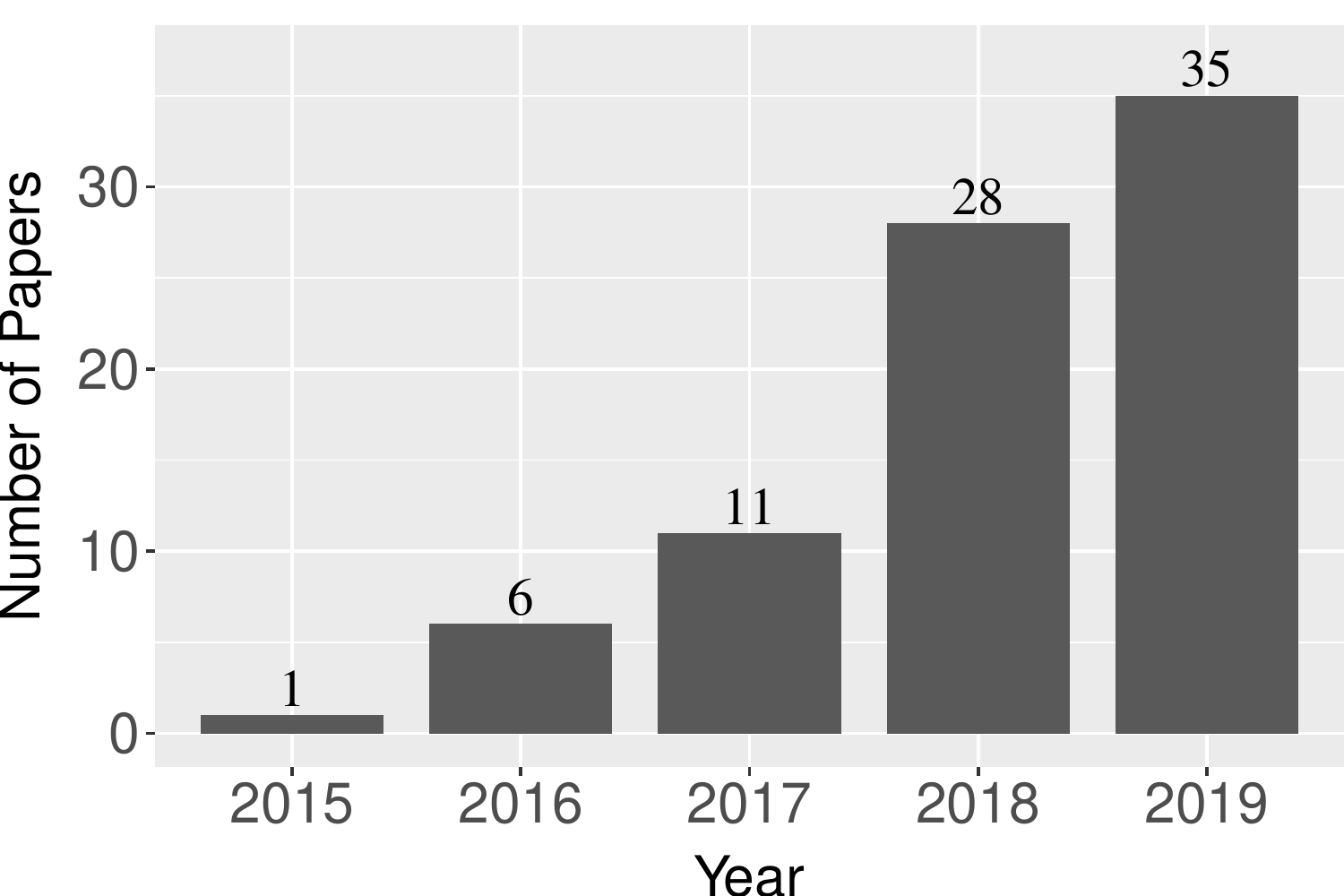}
  \vspace{-0.1cm}
  \caption{Papers by year}
  \label{fig:paper_by_year}
\end{figure}

The study selection resulted in 81 primary studies. We did not define
an initial publication date for the candidate papers.
Despite that, we found a single paper published
in 2015. All other papers are from subsequent
years, as illustrated in Figure~\ref{fig:paper_by_year}.


While we only consider nine months from 2019, we have more papers
published in 2019 than in 2018, which shows an increasing
interest in applying DL in
SE.



\begin{figure}[!t]
 \centering
 \includegraphics[width=.45\textwidth]{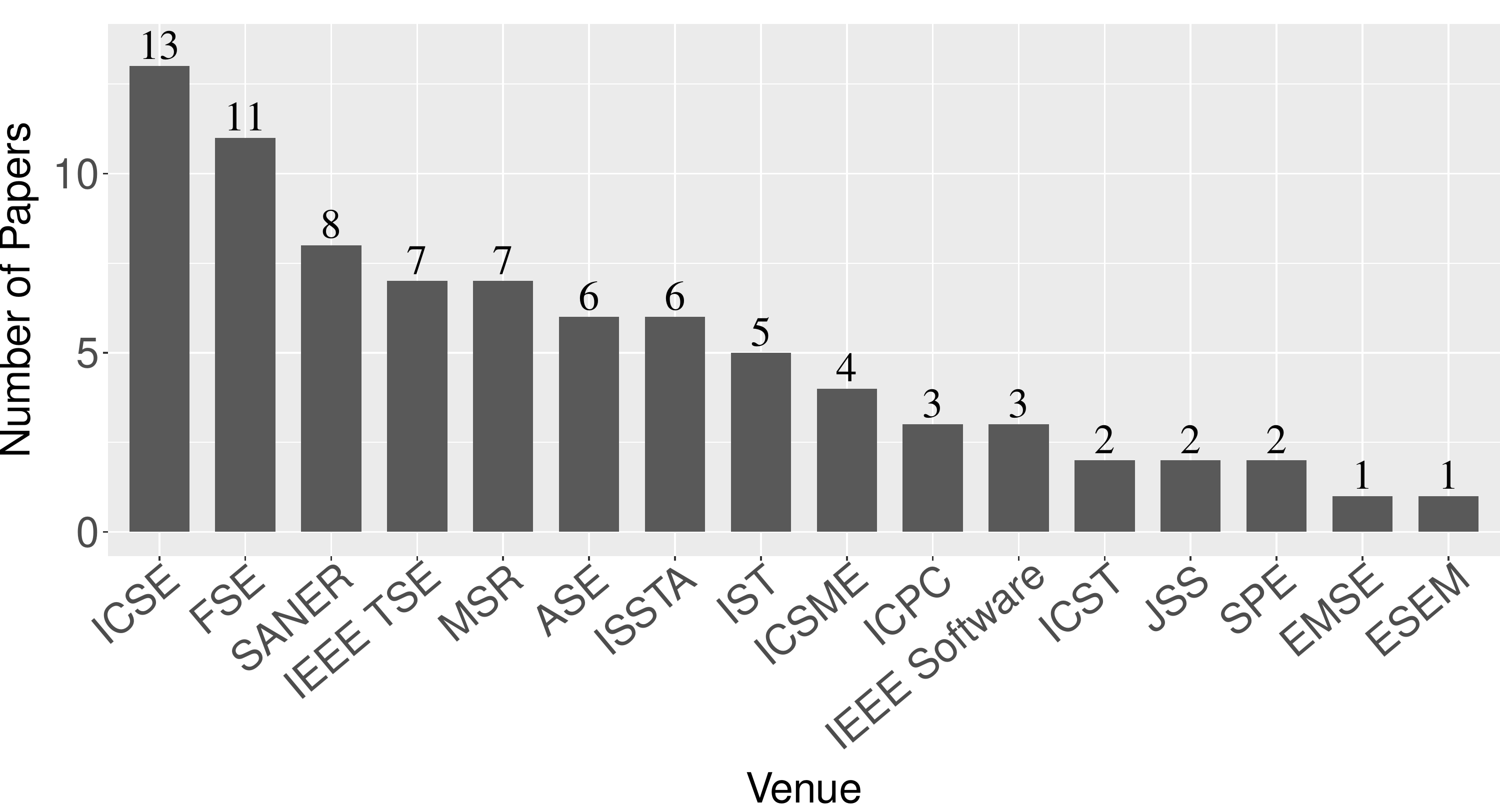}
  \vspace{-0.3cm}
 \caption{Papers by venue}
 \label{fig:paper_by_venues}
  \vspace{-0.3cm}
\end{figure}

Figure~\ref{fig:paper_by_venues} shows the number of papers by publication venue. Among our primary studies, we identified that 61 studies are from conferences (75.3\%) and 20 papers  from journals (24.7\%). ICSE and FSE concentrate most papers (24 papers or 29.6\%). IEEE TSE is the journal with the highest number of papers (7 papers, 8.6\%). We did not find papers about DL \& SE in nine venues: MODELS, SLPC, RE, FASE, ICSA, ACM TOSEM, SoSyM, SCP, and SQJ.

We found 12 papers (14.8\%)
with at least one author associated to industry.
Microsoft Research has the highest
number of papers (3 papers), followed
by Clova AI, Facebook, Grammatech,
Nvidia, Accenture, Fiat Chrysler, IBM,
and Codeplay (each one with a
single paper).

\subsection* {(RQ1) What SE problems are solved by DL?}

\begin{figure*}[!th]
  \centering
  \includegraphics[scale=1,width=0.95\textwidth]{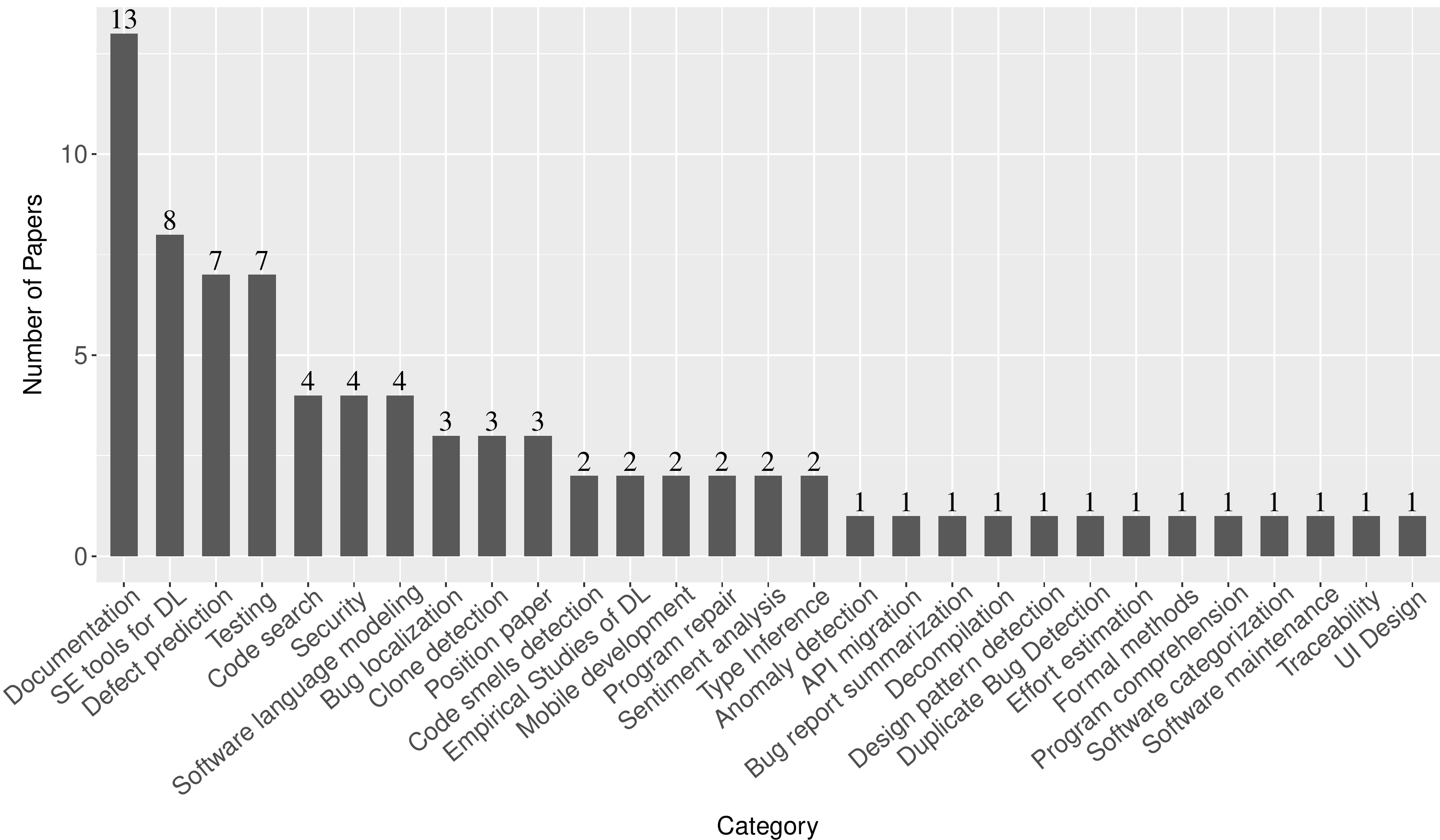}
  \caption{Papers by research problem}
  \label{fig:paper_by_category}
\end{figure*}

As illustrated in Figure~\ref{fig:paper_by_category}, we classified the papers in three principal groups: (1) papers that investigate the usage of SE tools and techniques in the development of DL-based systems; (2) papers that propose the usage of DL-based techniques to solve SE-related pro\-blems; and (3) position papers or tutorials. The following subsections describe the papers in each group.

\subsubsection{Using Software Engineering Techniques in DL-based Software}

We classified 10 papers in this category (12.3\%), including papers that adapt SE tools and techniques to DL-based software (8 papers) and  papers that describe empirical studies of DL-based software (2 papers). Papers that apply SE to DL are mostly focused on solving particular problems for testing DL-based software~\cite{p21,p40,p41,p46,p54,p56}. We also found papers that describe quantitative metrics to assess DL-based software~\cite{p45} and to support the deployment of DL-based software systems~\cite{p47}.
Finally, we found two empirical studies of DL-based software, both investigating the characteristics of the
bugs reported in such systems~\cite{p2,p10}.

\subsubsection{Using DL Techniques in SE Problems}

The usage of DL in SE concentrates in three main problems: documentation,
testing, and defect prediction. We provide more details in the following paragraphs:
\vspace{0.15cm}

\noindent{\bf Documentation:} This category has the highest number of papers (13 papers, 16\%). Seven papers study problems associated with Stack Overflow (SO)  questions and answers. For example, in a paper at ASE16, Xu~\textit{et al.}~\cite{p76} were one the first to use word embedding and CNN to link semantically related knowledge units (KU) at Stack Overflow 
They claim that traditional word representations 
miss many cases of potentially linkable KUs at Stack Overflow. Interes\-tingly, at FSE17, Fu~\textit{et al.}~\cite{p50} showed that a simple optimizer used together with the SVM algorithm can achieve similar results, but 84x faster than Xu's runtime performance. 
Next, at ESEM18, Xu~\textit{et al.}~\cite{p77} replicate the two previous studies using a large dataset.
They report that (1) the effectiveness of both approaches declined sharply in the new dataset; (2) despite that, SVM continues to outperform by a small margin DNN approaches in this new dataset; (3) however, both approaches are outperformed by another lightweight SVM method, called SimBow. 
Finally, at MSR18, Majumder~\textit{et al.}~\cite{p1} report significant runtime improvements to Fu~\textit{et al.}~\cite{p50} results. First, they cluster the data (with KMeans) and then build local models on the produced clusters.


At ASE16, Chen~\textit{et al.}~\cite{p65} rely on word embedding and CNN to translate Chinese queries to English before searching in Stack Overflow. The authors trained the CNN model with large amount of duplicate questions (0.3 million) in Stack Overflow and the corresponding Chinese translations of these questions translated by machine (Google Translate).

At MSR19, Wang~\textit{et al.}~\cite{p51} proposed DeepTip, a DL based approach that uses different CNN architectures to extract small, practical, and useful tips from developer answers.

At ICPC18, XingHu~\textit{et al.}~\cite{p24} proposed a new approach named DeepCom to generate code comments for Java methods automatically. The authors use a DNN that analyzes structural information of Java methods for better comments generation.

We also found two papers~\cite{p4,p8} about the automatic identification of source code fragments in videos. At ICSE19, Zhao~\textit{et al.}~\cite{p8} present a DL-based computer vision technique to recognize developer actions from programming screencasts automatically. The authors detect screen changes by image differencing and then recognize developer actions with a CNN model. At MSR18, Ott~\textit{et al.}~\cite{p4} also proposed a DL solution based on CNN and autoencoders to identify source code in images and videos. They identified the presence of typeset and handwritten source code in thousands of video images with 85.6\%-98.6\% of accuracy.

\vspace{0.1cm}



\noindent{\bf Testing:} We found seven papers (8.6\%) using DL in software testing, covering fuzzing tests~\cite{p20,p64,p81}, fault localization~\cite{p18,p39}, mutation testing~\cite{p11},
and testing of mobile apps~\cite{p14}. For example, the three papers related to fuzzing tests adopt the LSTM model to automate the generation of test cases. Specifically, at ASE17, Godefroid~\textit{et al.}~\cite{p64} introduce Learn\&fuzz, a first attempt to automatically ge\-nerate test data for the Microsoft Edge renderer. Their tool uses  a corpus of PDF files and LSTM network as the encoder and decoder of seq2seq model. The authors claim that grammar-based fuzzing is the most effective fuzzing technique despite the grammar being written manually, making it a laborious and time-consuming process. Later, at ISSTA18, Cummins~\textit{et al.}~\cite{p20} also claim that traditional approaches based on compiler fuzzing are time-consuming (e.g.,~they can take nine months to port from a generator to another) and require a huge effort. Then, they introduce DeepSmith~\cite{p20}, an approach that encodes a corpus of handwritten code and uses LSTM to speed up their random program generator (requiring less than a day to train). The authors also state that their framework requires low effort, detects a similar number of bugs compared to the state-of-the art, and reveal bugs not detected by traditional approaches. 


The last paper we analyzed concerning fuzzing tests is SeqFuzzer, by Zhao~\textit{et al.}~\cite{p81}, published at ICST19 . In this paper, the authors argue that the increasing diversity and complexity of industrial communication protocols is a major challenge for fuzz testing, especially when dealing with stateful protocols. In addition, they also assert that current approaches have a significant drawback, since the process to model the protocol for generating test cases is laborious and time-consuming. To tackle this problem, SeqFuzzer also uses a seq2seq model based on a LSTM network to detect protocol formats automatically and to deal with the temporal features of stateful protocols. The authors report that SeqFuzzer automatically generates fuzzing information with high receiving rates and has successfully identified various security vulnerabilities in the EtherCAT protocol.

In a paper published at SANER19, Zhang~\textit{et al.}~\cite{p18} proposed CNN-FL, an approach to improve fault localization effectiveness. The authors claim that preliminary works found promising insights to fault localization but the current results are still preliminary. They use execution data from test cases to train the CNN network and the results show that CNN-FL notably improves the fault localization effectiveness. 
After CNN-FL, at ISSTA19, Li~\textit{et al.}~\cite{p39} claim 
that traditional learning-to-rank algorithms are very effective on fault localization problems, but they fail to automatically select important features and discover new advanced ones. To tackle this problem, they propose DeepFL to automatically detect or gene\-rate complex features for fault location prediction. DeepFL uses LSTM and MLP networks on different feature dimensions, such as code-complexity metrics and mutation-based suspiciousness. The authors showed that DeepFL outperforms traditional algorithms on the same set of features (detects 50+ more faults within Top-1) and during prediction phase (1200x faster).

Finally, we found a paper related to mutation test~\cite{p11} and ano\-ther about testing of mobile apps~\cite{p14}. At ICST19, Mao~\textit{et al.}~\cite{p11} evaluate the effectiveness and efficiency of a predictive mutation testing technique under cross-project on a large dataset. They use 11 classification algorithms including traditional ML (e.g. Random Forest) and DL algorithms (MLP, CNN, and Cascading Forest). They report that Random Forest achieves similar performance than the evaluated DL algorithms. At ICSE17, Liu~\textit{et al.}~\cite{p14} train a RNN to learn from existing UI-level tests to generate textual inputs for mobile apps. The authors claim that despite the improvements on automated mobile testing, relevant text input values in a context remains a significant challenge, which makes hard the large-scale usage of automated testing approaches. They report that RNN models combined with a Word2Vec result in app-independent approach. 

\vspace{0.15cm}

\noindent{\bf Defect Prediction:}
We also found seven papers (8.6\%) that rely on DL for defect prediction. At ICSE16, Wang~\textit{et al.}~\cite{p15} argue that traditional
features used by defect predictors fail to capture semantic differences in programs. Then, they use a Deep Belief Network (DBN) to automatically learn semantic features from token vectors extracted from ASTs. For within-project defect prediction, they report an average improvement of 14.2\% in F-score. An extended version of their work appeared at TSE18. Among other improvements, the authors included four open-source commercial projects in their evaluation setup (e.g., google/guava and facebook/buck). 
At MSR19, Dam~\textit{et al.}~\cite{p67} also argue that traditional features do not 
capture the multiple levels of semantics in a source
code file. Then, they use LSTM to extract both syntactic and structural information from ASTs. They evaluate the model using real projects provided by an industrial partner (Samsung). 
They report a recall of 0.86 (23\% of improvement from Wang previous results), but at a lower precision. 
However, the authors claim
that recall is more important when predicting
defects, because the costs of missing defects
might be much higher than the ones of reporting
false defects.

At TSE18, Wen~\textit{et al.}~\cite{p55} use change sequences to predict defects, e.g., a conventional metric used by standard models might be the number of distinct developers who changed a particular file. Instead of using this single value, they use vectors with the name of developers responsible for each change performed in the file. The authors report F-measure of 0.657, which represents an improvement of 31.6\% over prediction models based on traditional metrics.
At MSR19, Hoang~\textit{et al.}~\cite{p42}
propose DeepJIT to automatically extract features from code changes and commit messages. The goal
is to use DL to extract features that represent the semantics of commits. The authors report improvements
of around 10\% when DeepJIT is compared
with a state-of-the-art benchmark. 

At IST18, Tong~\textit{et al.}~\cite{p82} argue that
Stacked Denoising Autoencoders (SDAEs) were never used in the field of software defect prediction. Finally, at IST19, Zhou~\textit{et al.}~\cite{p60} propose a new deep forest model 
to predict defects that relies on a cascade strategy to transform random forest classifiers into a layer-by-layer structure. 

\vspace{0.15cm}


\noindent{\bf Other research problems:} Other important research problems handled
using DL are code search~\cite{p23,p26,p33,p90},
security~\cite{p13,p49,p66,p72}, and software language 
modelling~\cite{p86,p12,p32,p44}. The next most investigated
research problem, with  three papers each,
are bug localization~\cite{p16,p38,p58} and 
clone detection~\cite{p17,p71,p30}.
We also found two papers on each of the
following problems: code smell detection~\cite{p63,p28},
mobile development~\cite{p35,p84}, program repair~\cite{p74,p83},
sentiment analysis~\cite{p61,p80}, and type inference~\cite{p34,p73}.

Finally, we found one paper related to the following problems: anomaly detection \cite{p78}, API migration \cite{p70}, bug report summarization \cite{p88}, decompilation \cite{p89}, design patterns detection \cite{p52}, duplicate bug detection \cite{p87}, effort estimation \cite{p5}, formal methods \cite{p37}, program comprehension \cite{p59}, software categorization \cite{p7}, software maintenance \cite{p43}, traceability \cite{p79}, and UI design \cite{p53}.


\subsubsection{Position Papers} 
We classified three papers (3.7\%) in this ca\-tegory, all published at IEEE Software.
They describe the challenges and opportunities of using DL in automotive software~\cite{p31,p91} or provide a short tutorial on machine learning and DL~\cite{p68}.


\subsection* {(RQ2) What DL techniques are used in SE problems?}

\begin{figure}[!b]
\vspace{-0.4cm}
  \centering
  \includegraphics[scale=1.0,width=0.46\textwidth]{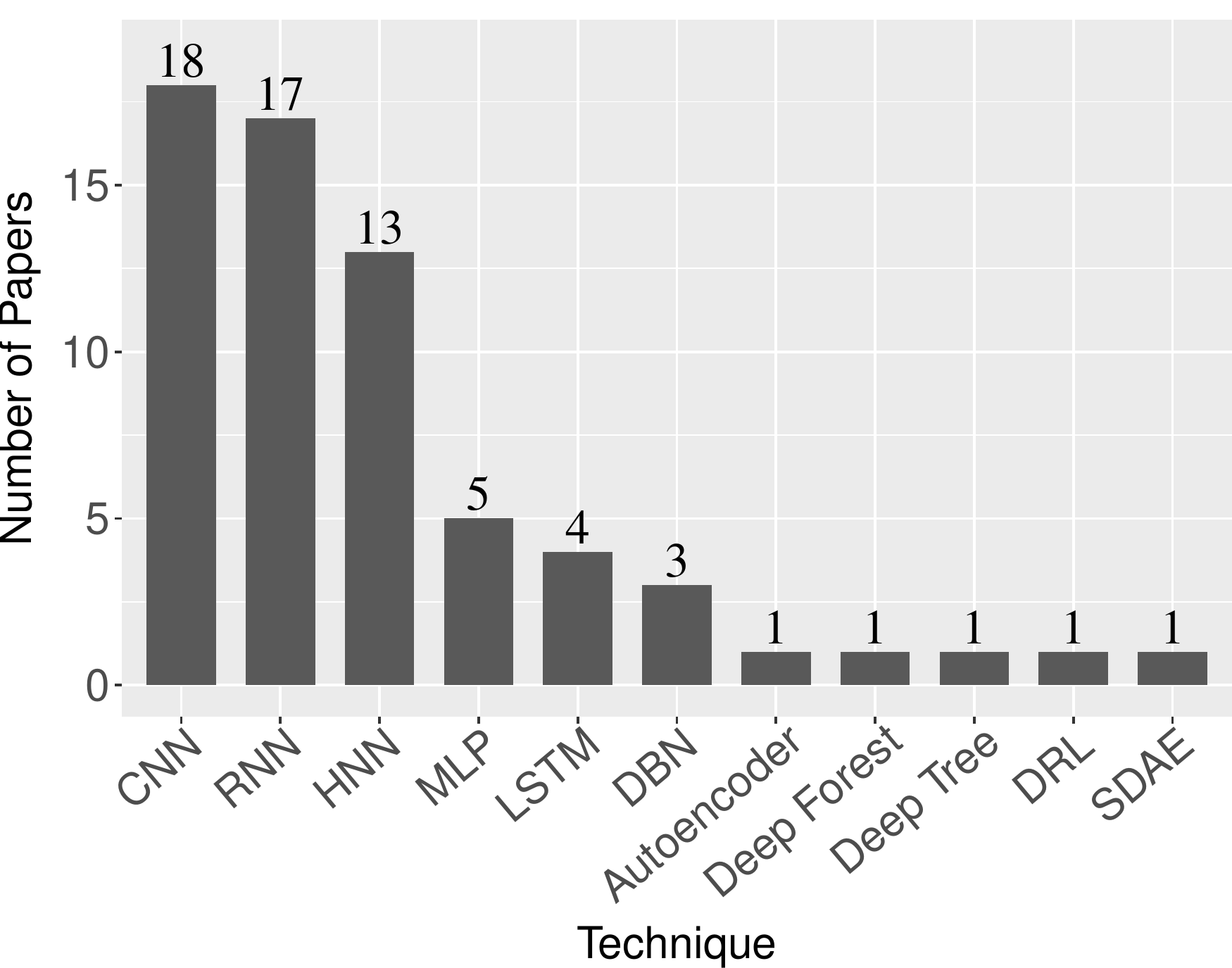}
  \caption{Papers by DL technique}
  \label{fig:paper_by_techniques}
\end{figure}

To answer this RQ, we identify the DL technique that each paper uses for supporting the SE research problem.
Figure~\ref{fig:paper_by_techniques} shows the most common DL techniques used by the analyzed papers. 
The most common techniques are CNN (18 papers, 22.2\%), RNN (17 papers, 20.9\%), and HNN (12 papers, 14.8\%). 


Table~\ref{tab:nn_problems}  shows the distribution of the
DL techniques by research problem. As we can observe, RNNs are used in all problems, except security and bug localization. Although CNN is used in more papers (18 papers), they have a focus on four problems (documentation, testing, bug localization, and clone detection).

\begin{table}[!t]
 \caption{Neural networks techniques by research problem}
 \centering 
  \begin{adjustbox}{width=.47\textwidth}
 \begin{tabular}{l c c c c c c c}
 \toprule
 & \multicolumn{6}{c}{\bf Neural Networks} \\
{\bf Problem} &
    {\bf CNN}&
    {\bf RNN}&
    {\bf HNN}&
    {\bf LSTM}&
    {\bf DBN}&
    {\bf MLP}\\
 \midrule
Documentation & $\bullet$ & $\bullet$ & $\bullet$ & \\
Defect prediction & & $\bullet$ & $\bullet$ & & $\bullet$ &\\
Testing & $\bullet$ & $\bullet$ & $\bullet$ & $\bullet$\\
Code search & &$\bullet$& & & $\bullet$ &\\
Security & & & $\bullet$ & $\bullet$ & & \\
Software language modeling & & $\bullet$ & & & & \\
Bug localization & $\bullet$ & & $\bullet$ & & & $\bullet$\\
Clone detection & $\bullet$ & $\bullet$ & & & & $\bullet$\\
 \bottomrule
\end{tabular}
 \end{adjustbox}
\label{tab:nn_problems}
\end{table}

\subsection* {(RQ3) How does DL compare with other machine learning techniques used in SE problems?}

The use of DL techniques to solve SE problems has increased signi\-ficantly. This popularity may lead researchers to adopt DL because they believe it will always be the best choice, or because of its fame. However, sometimes applying DL to a SE problem may not be the best approach and could even be a wrong choice considering its drawbacks. For example, Majumder~\textit{et al.}~\cite{p1} extend recent results for text mining Stack Overflow by clustering the dataset, then tuning every learner within each cluster. They obtained results over 500 times faster than DL (and over 900 times faster if they use all the cores on a standard laptop computer). Fakhoury~\textit{et al.}~\cite{p63} also compare traditional ML with DL techniques. The authors build and validate an oracle of around 1,700 instances and create binary classification models using traditional ML approaches and CNN. Their results show that traditional ML algorithms outperform deep neural networks in all evaluation metrics and resources (time and memory).


Mao~\textit{et al.}~\cite{p11} use ML and DL algorithms to evaluate the effectiveness and efficiency of a predictive mutation testing technique under cross-project dataset. The results show that Random Forest (ML algorithm) achieves nearly the same effectiveness for Predictive Mutation Testing compared to advanced DL models.





In some cases, adaptations in traditional machine learning techniques are sufficient to improve the results.  For example, Fu and Menzies~\cite{p50} use deep learning to find which questions in Stack Overflow can be linked together. They show that applying a very simple differential evolution optimizer to fine-tune SVM can achieve similar (and sometimes better) results. Their approach terminated in 10 minutes, i.e., 84 times faster than the deep learning method.

Zhou~\textit{et al.}~\cite{p60} investigate whether DL is better than traditional approaches in tag recommendation for software information sites. Their findings show that their model, with some engineering, can achieve better results than DL models. 

\section{Discussion}




As discussed in Section~\ref{sec:results}, the use of DL in SE problems has increased significantly. To analyze whether Deep Learning benefits the software engineering community, we focused on papers that use DL techniques to solve SE problems to identify the strengths and drawbacks of these techniques. 

\subsection{Strengths}


To understand why a significant number of papers are frequently adopting DL, we looked at papers that use DL techniques to solve SE-related problems aiming to identify their strengths. In what follows, we present the key strengths pointed out by the studies:

\vspace{0.2cm}

\noindent{\bf Best results with unstructured data:} 
most ML algorithms face difficulties to analyze unstructured data like pictures, pdf files, audios, and more. In contrast, it is possible to use different data formats to train DL algorithms and obtain insights that are relevant to the training purpose. For example, Zhou~\textit{et al.}~\cite{p8} recognize develo\-pers actions from programming screencasts. Using programming screencasts from Youtube, they demonstrate the usefulness of their technique in a practical scenario of action-aware extraction of key-code frames in developers’ work. Ott~\textit{et al.}~\cite{p4} use a similar DL approach to identifying source code in images and videos. 
\vspace{0.15cm}

\noindent{\bf No need for Feature Engineering (FE):} 
FE is an essential task in ML, once it improves model accuracy. However, it often requires domain knowledge in a specific problem. The main advantage of DL techniques is their ability to apply FE by themselves. DL algorithms scan the data to find correlated features and combine them through multi-layers to enable faster learning. According to Zhou~\textit{et al.}~\cite{p8}, e.g., recognizing developers' actions from programming screencasts is a challenging task due to the diversity of developer actions, working environments, and programming languages. Thus, they use the CNN’s ability to automatically extract image features from screen changes resulting from developer actions. 

\vspace{0.15cm}
\noindent{\bf No need for labeling data:} labeling process gets a set of unlabeled data and adds meaningful tags for each piece of these data. For example, labels may indicate whether an image contains a dog or a cat, what is the sentiment of a post, or what is the action performed in a video. 
ML requires labeled data for training, which may be expensive since these algorithms usually require thousands of data. By contrast, 
DL supports unsupervised learning techniques that allow the learning with no guidelines. For example, Dam~\textit{et al.}~\cite{p13} use LSTM to detect the most likely locations of code vulnerabilities in an unlabeled code base. According to the authors, LSTM offers a powerful representation of source code and can automatically learn both syntactic and semantic features that represent long-term source code dependencies. 
\vspace{0.15cm}

\noindent{\bf Best results with sequence prediction:} RNNs are designed to receive historical sequence data and predict the next output value in the sequence. Unlike traditional ML techniques, RNNs can embed sequential inputs, such as sentences in their internal memory. This step allows them to achieve better results for tasks that deal with sequential data, e.g., text and speech~\cite{p50}. 
\vspace{0.15cm}

\noindent{\bf High-quality results:} once trained correctly, DL techniques can perform thousands of repetitive tasks with better performance compared to traditional ML techniques~\cite{p55}. 

\vspace{0.15cm}








We summarize these strengths and the papers that cited each of them in Table~\ref{tab:strengths}.
\begin{table}[!t]
 \caption{Strengths}
 \vspace{-0.3cm}
 \centering 
  \begin{adjustbox}{width=.47\textwidth}
\begin{tabular}{lrp{40mm}} \toprule
\textbf{Strength} & \textbf{\#} & \textbf{References}\\ \midrule
Best results with unstructured data & 5 & \cite{p53}, \cite{p4}, \cite{p8}, \cite{p63},  \cite{p89}\\
No need for feature engineering & 12 & \cite{p4}, \cite{p7}, \cite{p8}, \cite{p13}, \cite{p24}, \cite{p52}, \cite{p42}, \cite{p55}, \cite{p36}, \cite{p67}, \cite{p66}, \cite{p59}\\
No need for labeling data & 6 & \cite{p23}, \cite{p13}, \cite{p80}, \cite{p88}, \cite{p36}, \cite{p67}\\
Best results with sequential data & 6 & \cite{p14}, \cite{p34}, \cite{p90}, \cite{p50}, \cite{p89}, \cite{p55}\\
High-quality Results & 10 & \cite{p5}, \cite{p18}, \cite{p24}, \cite{p80}, \cite{p86}, \cite{p87}, \cite{p88}, \cite{p17}, \cite{p26}, \cite{p28}\\

\bottomrule 
\end{tabular}
 \end{adjustbox}
\label{tab:strengths}
\vspace{-0.2cm}
\end{table}
\subsection{Drawbacks}

Despite their benefits, DL-based techniques also have some drawbacks (which are summarized in Table~\ref{tab:drawbacks}). 

\vspace{0.15cm}
\noindent{\bf Computational cost:} the process of training is the most challen\-ging part of using DL-based techniques, and by far the most time consuming since DL algorithms learn progressively. Consequently, DL-based techniques require high-performance hardware, such as hundreds of machines equipped with expensive multi-core GPUs. According to Fu and Menzies~\cite{p50}, even with advanced hardware, training DL models requires from hours to weeks. We found refe\-rences to high computational cost in at least 22 papers. 
\vspace{0.15cm}

\noindent\textbf{Size and quality of the dataset:} DL algorithms require extensive amounts of data for training. While companies like Google, Facebook, and Microsoft have abundant data, most researchers do not have access to a massive amount of data. For example, the effectiveness of DeepSims, a DL approach proposed by Zhao and Huang~\cite{p44}, is limited by the size and quality of the training dataset. The authors state that \noindent{\em``building such a large and representative dataset is challenging".}
\vspace{0.15cm}

\noindent\textbf{Replication:} once trained with data, DL models are very efficient in formulating an adequate solution to a particular problem. However, they are unable to do so for a similar problem and require retraining. Moreover, reproducing DL results is also a significant problem~\cite{p1}. Majumder~\cite{p1} state that: \noindent{\em``it is not yet common practice for deep learning researchers to share their implementations and data, where a tiny difference may lead to a huge difference in the results".}
\vspace{0.15cm}



\noindent\textbf{Number of parameters:}  DL techniques are multi-layered neural networks. The number of parameters grows with the number of layers, which might lead to overfitting, a tendency to “learn” characteristics of the training data that does not generalize to the population as a whole. At least 11 papers reference this problem. 

\begin{table}[!ht]
 \caption{Drawbacks}
 \vspace{-0.3cm}
 \centering 
  \begin{adjustbox}{width=.47\textwidth}
\begin{tabular}{lrp{60mm}} \toprule
\textbf{Drawback} & \textbf{\#} & \textbf{References}\\ \midrule
Computational Cost & 22 & \cite{p1},  \cite{p4}, \cite{p23}, \cite{p50}, \cite{p90}, \cite{p12}, \cite{p18}, \cite{p26}, \cite{p34}, \cite{p37}, \cite{p53}, \cite{p87}, \cite{p89}, \cite{p16}, \cite{p63}, \cite{p42}, \cite{p60}, \cite{p67}, \cite{p30}, \cite{p28}, \cite{p59}, \cite{p66} \\ 
Dataset size and quality & 7 & \cite{p44}, \cite{p28}, \cite{p63}, \cite{p55}, \cite{p60}, \cite{p66}, \cite{p59}\\
Replication & 3 & \cite{p1}, \cite{p63}, \cite{p55}\\
Number of parameters & 11 & \cite{p1}, \cite{p5}, \cite{p12}, \cite{p52}, \cite{p15}, \cite{p28}, \cite{p63}, \cite{p60}, \cite{p61}, \cite{p67}, \cite{p72}\\
Implementation Difficulty & 4 & \cite{p1}, \cite{p63}, \cite{p59}, \cite{p60}\\
\bottomrule 
\end{tabular}
 \end{adjustbox}
\label{tab:drawbacks}
\vspace{-0.4cm}
\end{table}

\section{Conclusion}
\label{sec:conclusion}

In this work, we analyzed 81 recent papers that apply DL techniques to SE problems or vice-versa. Our main findings are as follows: i) DL is gaining momentum among SE researchers. For example, 35 papers (43.2\%) are from 2019 and only one paper from 2015; ii) The top-3 research problems tackled by the analyzed papers are documentation (13 papers), defect prediction (7 papers), and testing (7 papers), and iii) The most common neural network type used in the analyzed papers is CNN and RNN.

The list of papers and the data
analyzed in this work are available at: \url{https://doi.org/10.5281/zenodo.4302713}

\section*{Acknowledgments}

\noindent Our research is supported by FAPEMIG and CNPq.

\small
\balance
\bibliographystyle{IEEEtran}
\bibliography{bibfile}

\end{document}